\documentclass[12pt]{article}
\newenvironment{wileykeywords}{\textsf{Keywords:}\hspace{\stretch{1}}}{\hspace{\stretch{1}}\rule{1ex}{1ex}}

\usepackage{amsmath,amssymb}
\usepackage{graphicx}
\usepackage{color}
\usepackage{dcolumn}
\usepackage{bm}
\usepackage[numbers,super,comma,sort&compress]{natbib}
\usepackage{braket}

\usepackage[usenames,dvipsnames,svgnames,table]{xcolor}
\newcommand{\orbkit}{\textsc{orbkit}}

\definecolor{background-color}{gray}{0.98}

\title{\orbkit{} -- A Modular Python Toolbox \\for Cross-Platform Post-Processing\\ of Quantum Chemical 
Wavefunction Data}
\author{
Gunter Hermann\thanks{Institut f{\"u}r Chemie und Biochemie, Freie 
Universit{\"a}t Berlin, Takustra{\ss}e 3, 14195 Berlin, Germany} \thanks{These 
authors contributed equally to this work.}, 
Vincent Pohl\footnotemark[1] \footnotemark[2], 
Jean Christophe Tremblay\footnotemark[1],\\
Beate Paulus\footnotemark[1],
Hans-Christian Hege\thanks{Department of Visual Data Analysis, 
Zuse Institute Berlin, Takustra{\ss}e 7, 14195 Berlin, Germany}, 
Axel Schild \thanks{Max-Planck-Institut f{\"u}r 
Mikrostrukturphysik, Weinberg 2, 06120 Halle, Germany}}

\begin{document}

\maketitle

\begin{abstract}
\orbkit{} is a toolbox for post-processing electronic structure calculations based 
on a highly modular and portable Python architecture. 
The program allows computing a multitude of electronic properties of molecular 
systems on arbitrary spatial grids from the basis set representation of its 
electronic wavefunction, as well as several grid-independent properties. 
The required data can be extracted directly from the standard output of a large 
number of quantum chemical programs. 
\orbkit{} can be used as a standalone program to determine standard quantities, 
for example, the electron density, molecular orbitals, and derivatives thereof. 
The cornerstone of \orbkit{} is its modular structure. 
The existing basic functions can be arranged in an individual way and can be easily 
extended by user-written modules to determine any other desired quantities. 
\orbkit{} offers multiple output formats that can be processed by common visualization 
tools (VMD, Molden, etc.). 
Additionally, \orbkit{} possesses routines to order molecular orbitals computed 
at different nuclear configurations according to their electronic character 
and to interpolate the wavefunction between these configurations. 
The program is open-source under GNU-LGPLv3 license and freely available at 
http://sourceforge.net/projects/orbkit/. 
This article provides an overview of \orbkit{} with particular focus on its 
capabilities and applicability, and includes several example calculations.
\end{abstract}

\begin{wileykeywords}
quantum chemical calculation, electronic structure, molecular visualization, electron density, 
grid representation of one-electron quantities, molecular orbital ordering
\end{wileykeywords}

\clearpage


\begin{figure}[h]
\centering
\colorbox{background-color}{
\fbox{
\begin{minipage}{1.0\textwidth}
\includegraphics[width=50mm,height=50mm]{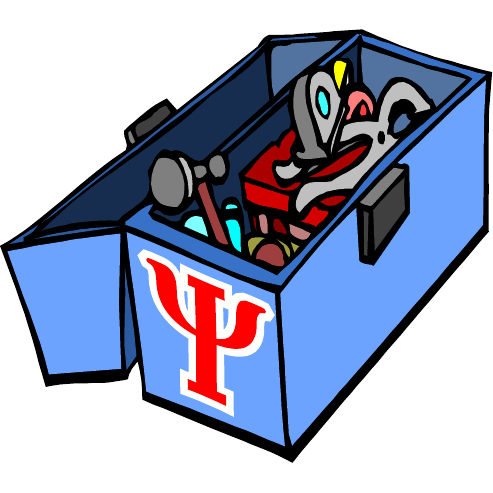}\\
\orbkit{} is an open-source toolbox for post-processing electronic structure calculations. 
Based on a highly modular and portable Python architecture, it comes both as a 
standalone program and a function library.
The program allows computing electronic properties of molecular 
systems on arbitrary spatial grids from the output of standard quantum chemistry
programs.
\end{minipage}
}}
\end{figure}

  \makeatletter
  \renewcommand\@biblabel[1]{#1.}
  \makeatother

\bibliographystyle{apsrev}

\renewcommand{\baselinestretch}{1.5}
\normalsize

\clearpage

\section*{\sffamily \Large Introduction}

In today's computational and theoretical chemistry, quantum chemical methods (or electronic structure methods) are routinely applied for the investigation of molecular systems.
Modern quantum chemical programs are characterized by their general applicability, increasing functionality, and high efficiency due to methodological and numerical progress in the field.
There exists a wide range of such program packages, covering different levels of theory and offering assorted features.
The spectrum extends from open source packages, e.g., GAMESS-US\cite{gamess}, PSI4\cite{psi4}, or Tonto\cite{tonto}, over commercial closed source programs such as Gaussian\cite{gaussian}, Molpro\cite{molpro}, Turbomole\cite{turbomole}, or Q-Chem\cite{qchem}, to software freely available for academic usage such as ORCA\cite{orca} or NWChem\cite{nwchem}.
Depending on the methodological requirements of a quantum chemical problem, the user has to deal with a multitude of differently formatted input and output data. 
In this context, projects such as the Atomic Simulation Environment (ASE)\cite{ASE}, the Basis Set Exchange library\cite{bse1,bse2}, cclib\cite{cclib}, and OpenBabel\cite{openbabel} contributed tremendous standardization efforts.

Despite this software diversity, most quantum chemical programs dealing with molecules share the same basic approach to the solution of the time-independent molecular Schr\"odinger equation: They solve it for clamped nuclei, and they use an atom-centered Gaussian basis set to represent the electronic wavefunction at the selected nuclear configuration.
A typical output of a quantum chemical calculation contains not only the energy and other relevant properties of the molecular system, but also the expansion coefficients of the electronic wavefunction in the selected basis set.
The latter allow the calculation of additional quantities for the system characterization. 
Conceivable quantities include those based on the reconstructed electronic wavefunction, e.g., the electron density, and others quantities, such as molecular orbitals (MO), that are constructed from various combinations of the basis set and the expansion coefficients.
These quantities are typically represented on a grid in the configuration space of one electron, which facilitates their analysis and enables their visualization.
The necessary post-processing tools for the analysis are sparsely implemented in most of the quantum chemical program packages. 
Additionally, there is usually no opportunity to adjust the post-processing parameters, e.g., grid parameters, or to request further quantities after finishing the electronic structure calculation.

To overcome this problem, two strategies can be pursued: the modification or extension of a quantum chemical program package, or the usage of standalone post-processing programs.
The first approach is practicable only for open source and well-documented programs; additionally it is a formidable, time-consuming work to understand, adapt and extend the respective source code.
For the second approach, a handful of specialized tools are available offering diverse functionalities.
An easy visualization of the molecular structure, the MOs, the electron density, etc., based on the output of a quantum chemical program, can be carried out with programs such as Molden\cite{molden} or Avogadro\cite{avogadro}.
To calculate properties from the electronic wavefunction (i.e., from the basis set used and the coefficients obtained in the electronic structure calculation) the programs Checkden\cite{checkden1,checkden2}, DGrid\cite{dgrid}, Multiwfn\cite{multiwfn}, or DensToolKit\cite{denstoolkit} are well-suited and provide an impressive number of features.
However, if the desired feature is not already available, extending these codes may become prohibitively difficult.

For such problems, we have developed the Python toolbox \orbkit{}, which meets all common requirements of post-processing electronic wavefunctions.
\orbkit{} stands out by its broad applicability in terms of post-processing electronic structure data.
It offers similar features such as Checkden, DGrid, or Multiwfn, and it can be employed as a standalone program for investigating in detail the characteristics of a molecular system and for visualizing important position space quantities.
Its modular design allows the user to combine the individual functions in any manner.
Furthermore, \orbkit{} also comes with a library and application programming interface. This can be used to create new programs, without in-depth knowledge of the internal structure. Additionally, the library can be extended by user-written functions.
Programming is greatly facilitated by using Python as major language with its user-friendly syntax and large number of function libraries.
Consequently, also non-standard or new problems can be quickly solved by adding new user-written functions to the standalone program \orbkit{}, or by combining existing functions and new functions in a user-written program.
In this article, we want to present the capabilities and several selected applications of \orbkit{}.

The paper is structured as follows. Sec.\ ``Methodology'' briefly introduces the theoretical background and Sec.\ ``Program'' describes the structure and main aspects of \orbkit{}. Then, we present several ``Practical Applications'' that illustrate the features of \orbkit{}, followed by a conclusion.

\section*{\sffamily \Large Methodology}

In this section, we present the main functions implemented in \orbkit{} that are necessary to post-process results from electronic structure calculations. 
Quantities that can be constructed from these fundamental components and that are included in \orbkit{} are not listed here but presented in Sec.\ ``Program". All theoretical aspects and quantities considered in Sec.\ ``Practical Applications" will be briefly discussed there, in the respective subsection. We use atomic units throughout the article.

\subsection*{\sffamily \large The Electronic Wavefunction}

For the solution of the time-independent molecular Schr{\"o}dinger equation for clamped nuclei, most molecular quantum chemistry methods introduce localized one-electron basis set functions to expand the many-body electronic wavefunction.
Accordingly, a standard output of a quantum chemical program provides the data to reconstruct the electronic wavefunction, i.e., the expansion coefficients of the wavefunction, the selected atom-centered basis set (the atomic orbitals), and the nuclear configuration (position and type of the nuclei).
Gaussian-type orbitals are by far the most commonly used atom-centered basis sets, and they can be handled with \orbkit{}.

In general, the many-body electronic wavefunction is arranged in the form of a single Slater determinant or a linear combination of multiple Slater determinants. These are defined as antisymmetrized products of $N$ one-electron functions which correspond to orthonormal MOs.
In the MO-LCAO (Molecular Orbital - Linear Combination of Atomic orbitals) ansatz, each  MO $\varphi_{a}$ can be reconstructed by the linear expansion of a finite set of contracted Gaussian basis functions\cite{jensen}
\begin{eqnarray}\label{molcao}
\varphi_{a} \left(\mathbf{r}\right) = \sum_{A}^{N_{\rm{A}}}\sum_{i}^{N_{\rm{AO}}} C_{ia} \psi_{i}\left(\mathbf{r}-\mathbf{R}_{A}\right),
\end{eqnarray}
where $C_{ia}$ is the $i$th MO coefficient for the MO $a$, $\psi_{i}$ is the respective atomic orbital centered at atom $A$, $\mathbf{r}$ are the Cartesian coordinates of one electron, $\mathbf{R}_{A}$ denotes the spatial coordinates of nucleus $A$, $N_{\rm{AO}}$ represents the number of atomic orbitals, and $N_{\rm{A}}$ is the number of atoms.

The atomic orbitals correspond to the real-valued contracted Gaussian basis functions which are defined by the linear combination of primitive Gaussian functions\cite{huzinaga84}
\begin{eqnarray}\label{cgto}
\psi_{i} \left(\mathbf{r_A}\right) = \sum^{L}_{p} d_{p} g_{p}\left(\mathbf{r_A},\alpha_{p},l_p,m_p,n_p \right),
\end{eqnarray}
where $d_{p}$ labels the contraction coefficients, $L$ is the length of the contraction, and $\mathbf{r_A}=\mathbf{r}-\mathbf{R_A}$ is the position
vector of an electron relative to the nucleus $A$.

A primitive Cartesian Gaussian function $g_{p}$ has the form\cite{huzinaga84}
\begin{eqnarray}\label{gto}
g_{p}\left(\mathbf{r},\alpha_{p},l_p,m_p,n_p\right) = N_{p}\left(\alpha_{p},l_{p},m_{p},n_{p}\right) x^{l_p} y^{m_p} z^{n_p} \exp \left(-\alpha_{p} r^2\right),
\end{eqnarray}
where $x$, $y$, and $z$  are Cartesian coordinates, $r=\sqrt{x^2+y^2+z^2}$ is the magnitude of $\mathbf{r}$, $\alpha_{p}$ labels the Gaussian orbital exponents, and $l_p$, $m_p$, and $n_p$ are declared as exponents whose sum determines the angular momentum and, thus, the type of the orbital, e.g., $l = l_p+m_p+n_p = 0$ for an s-orbital.

The normalization constant $N_{p}$ is given by\cite{huzinaga84}
\begin{eqnarray}
N_{p}\left(\alpha_{p},l_{p},m_{p},n_{p}\right) = \sqrt{\left(\frac{2\alpha_{p}}{\pi}\right)^{\frac{3}{2}}\frac{(4\alpha_{p})^{l_{p}+m_{p}+n_{p}}}{(2l_{p}-1)!! \, (2m_{p}-1)!! \, (2n_{p}-1)!!}}.
\end{eqnarray}
In addition to the Cartesian Gaussians, \orbkit{} can process spherical harmonic Gaussians by using the transformation described by Schlegel and Frisch.\cite{schlegel_trans}

As a position space one-electron quantity, the experimentally observable electron density can provide further insights for the analysis of electronic and chemical characteristics of the system, for instance, bonding properties. These can be studied in various ways, for example, in partial charge analysis methods such as the Voronoi Deformation Density (VDD).\cite{vdd}
For a single Slater determinant ansatz, the one-electron density reads
\begin{eqnarray}\label{el_dens}
\rho_{e}\left(\mathbf{r}\right) =  \sum_{a}^{N_{\rm occ}} n_{a} \left| \varphi_{a}\left(\mathbf{r}\right) \right|^2,
\end{eqnarray}
where $N_{\rm{occ}}$ is the number of singly or doubly occupied MOs with the respective occupation number $n_a$.
For multi-determinant wavefunctions, as obtained from configuration interaction methods or coupled cluster methods, 
the one-electron density can be constructed, for instance, by using the Slater-Condon rules or by converting the wavefunction into a 
single Slater determinant representation built from natural orbitals. 
These orbitals possess non-integer occupation numbers $n_a$ between zero and two.\cite{jensen} 
The default version of \orbkit{} supports all single-determinant wavefunctions.
Hence, it can directly be used for the results of a Hartree-Fock or Density Functional Theory calculation as well as of a Post-Hartree Fock calculation in natural orbital representation.
The evaluation of quantities directly from a multi-determinant wavefunction can be accomplished by using the Slater-Condon rules mentioned above.
For this purpose, it is straightforward to extend \orbkit{} by self-written modules.

\subsection*{\sffamily \large Analytical Derivatives and Integrals}

Further basic components, which can be derived from an electronic structure calculation and are relevant for the determination of other post-processing quantities, include analytical derivatives and integrals of the basis functions, MOs, and of the electron density.

One of these components is the gradient of the electron density with respect to the electronic coordinates, which has the general form
\begin{eqnarray}
\vec{\nabla}\cdot \rho_{e}\left(\mathbf{r}\right) =  2 \sum_{a}^{N_{\rm occ}} n_{a} \left( \varphi_{a}\left(\mathbf{r}\right) \vec{\nabla}\cdot \varphi_{a}\left(\mathbf{r}\right) \right)
\end{eqnarray}
and is constructed from the analytical gradients of the primitive Gaussian functions within the MO-LCAO ansatz
\begin{eqnarray}
\vec{\nabla}\cdot g_{p}\left(\alpha_{p},l_{p},m_{p},n_{p},\vec{r}\right)&=&\left(\begin{array}{c}
\frac{\partial}{\partial x}\\
\frac{\partial}{\partial y}\\
\frac{\partial}{\partial z}
\end{array}\right)g_{p}\left(\alpha_{p},l_{p},m_{p},n_{p},\vec{r}\right)\\&=&\left(\begin{array}{ccc}
l_{p}x^{-1} & - & 2\alpha_{p}x\\
m_{p}y^{-1} & - & 2\alpha_{p}y\\
n_{p}z^{-1} & - & 2\alpha_{p}z
\end{array}\right)g_{p}\left(\alpha_{p},l_{p},m_{p},n_{p},\vec{r}\right).
\end{eqnarray}
Note that the contraction coefficients $d_{p}$ from Eq. \ref{cgto} and the MO expansion coefficients $C_{ia}$ from Eq. \ref{molcao} are independent of the electronic coordinates and thus not affected by the derivative operator. The gradients can be used to calculate, e.g., the transition electronic flux density, as shown in an application below.

Closely related to the gradient is the Laplacian of the electron density, which is defined as
\begin{eqnarray}\label{laplacian}
\nabla^2 \rho_{e}\left(\mathbf{r}\right) = \left(\frac{\partial^2}{\partial x^2} +
\frac{\partial^2}{\partial y^2}+
\frac{\partial^2}{\partial z^2}\right)
\rho_{e}\left(\mathbf{r}\right).
\end{eqnarray}
By revealing information about the local depletion and concentration of the electron density, the Laplacian plays a key role in the specification of bonding properties, and therefore, it is also used in the Atoms in Molecules (AIM) theory.\cite{aim1,aim2}\\
To complete the set of essential functions incorporated in \orbkit{}, we introduce the MO overlap matrix 
\begin{eqnarray}\label{moom}
\Braket{\varphi_{a}|\varphi_{b}} & = & \Braket{\sum_{i}^{N_{\rm{AO}}}C_{ia}\psi_{i}|\sum_{j}^{N_{\rm{AO}}}C_{jb}\psi_{j}}\nonumber\\
 & = & \sum_{i}^{N_{\rm{AO}}}\sum_{j}^{N_{\rm{AO}}}C_{ia}C_{jb}\Braket{\psi_{i}|\psi_{j}}
\end{eqnarray}
and the atomic orbital overlap matrix
\begin{eqnarray}
\Braket{\psi_{i}|\psi_{j}} & = & \Braket{\sum_{p}^{L}d_{pi}g_{p}|\sum_{q}^{L}d_{qj}g_{q}}\nonumber\\
& = & \sum_{p}^{L}\sum_{q}^{L}d_{pi}d_{qj}\Braket{g_{p}|g_{q}}.
\end{eqnarray}
Here, $\Braket{g_{p}|g_{q}}$ is calculated based on the article of H{\^o} and  Hern{\'a}ndez-P{\'e}rez.\cite{gaussian_int} Possible quantities that can be derived from these overlap matrices involve, for example, Mulliken and L{\"o}wdin atomic populations or total and transition dipole moments.\cite{jensen}

\section*{\sffamily \Large Program}

For the development of \orbkit{}, we pursued the goal to make it practically useful for a large group of users, ranging from end-users who are interested in a simple and straightforward calculation of standard quantities, via university teachers who want to illustrate the computational ideas of quantum chemistry, to developers looking for a toolbox that provides core functions to build upon. To this end, we made a number of design decisions:
First, we chose Python as a programming language because of its user-friendliness, its vast amount of standard libraries, and its cross-platform portability.
Furthermore, we tried to retain a readily comprehensible modular structure, i.e., we implemented a broad set of functions that can be separately called in a user-assembled driver program.
This design facilitates the execution of the single components of \orbkit{} and opens up the opportunity to implement self-written features in combination with the already existing functions.
Besides, there is a standalone version of \orbkit{} which can calculate a selected number of quantities, such as the one-electron density or the MOs on a user-defined rectilinear grid.
Its simple handling allows quickly getting started with \orbkit{}.
In general, we attempt to ensure a universal applicability that comprises the readability of standard quantum chemistry programs, the writing of output files which are easy to handle, and an adequate number of standard quantities.
Tab. \ref{tab:features} gives an overview of the possible input and output file formats and lists the computable quantities.
 
Concerning the efficiency of \orbkit{}, we use the highly scalable NumPy\cite{numpy} and SciPy\cite{scipy} Python libraries for processing large, multi-dimensional arrays. 
Moreover, computationally expensive parts are implemented in C++ using the Python package weave\cite{scipy} and can be run on multiple processors by using the Python package multiprocessing. Position space quantities are then calculated by dividing the grid into slices and distributing them on the requested number of processors, thus offering linearly scaling parallelization.

To understand how to use \orbkit{}, it is recommended to read the detailed documentation and to work through the example applications in this article or in the \orbkit{} example package. The documentation also contains function references for advanced usage.

\begin{table}[htbp]
 \centering {\footnotesize
 \begin{tabular}{l|l}
 \hline
  Input                               &  Output     \\  \hline 
  Molden files                        &  HDF5 files \\
  AOMix files                         &  Gaussian cube files  \\
  GAMESS-US output files              &  VMD script files  \\
  Gaussian log-files                  &  ZIBAmiraMesh files and network files  \\
  Gaussian formatted checkpoint files      &  Mayavi visualization \\
  cclib library                              &  XYZ and PDB files \\  \hline \hline
  Standard Quantities                        &   Additional Feature               \\  \hline 
  Electron densities                         &   Input of real-space grid as regular grid or as point set           \\
  Atomic and Molecular orbitals              &   Order molecular orbitals of different nuclear structures  \\
  Orbital derivatives                        &   Interpolate between different nuclear structures     \\
  Gross atomic densities                     &   Symmetry transformations of the Cartesian grid    \\
  Molecular orbital transition flux densities&   Center grid around specified nuclei \\
  Total dipole moments                        &     \\
  Mulliken and L{\"o}wdin charges            &        \\ \hline
 \end{tabular} }
  \caption{Available input and output formats as well as computable quantities and other features of \orbkit{}.}
 \label{tab:features}
\end{table}

\subsection*{\sffamily \large Input and Output}

In general, \orbkit{} requires as main input the data of a single determinant wavefunction. 
To this end, it extracts the expansion coefficients of the MOs, the selected atomic basis set, and the specification of the molecular structure from the output of a quantum chemical calculation.
The data files of almost all major quantum chemical programs can be handled with \orbkit{} (cf. Tab. \ref{tab:features}).
Besides output files, such as Gaussian log-files, the Molden file format\cite{mofile} is our main input format. 
This file format can be both directly written by some programs, such as Molpro\cite{molpro}, and transformed with Molden\cite{molden} from output files of other program packages, such as GAMESS-US.
In addition, there exists an interface to the library cclib\cite{cclib}, which is a package-independent platform for parsing and extracting information of several computational chemistry programs.
With this extension, many further file formats can be read.
However, it is also straightforward to read the output of any other electronic structure program with a self-written Python routine and transfer it into the specific \orbkit{} data structure. The data formats used by \orbkit{} are described in detail in the documentation.

Based on the extracted data, \orbkit{} can compute all standard position space functions (cf. Tab. \ref{tab:features}) on an equidistant zero- to three-dimensional Cartesian grid with user-defined grid parameters.
Besides, the calculation can be performed on a list of $(x,y,z)$ coordinates which we call vector grids. This includes equidistant spherical coordinates, random grids, or user-defined arbitrary point sets.
It is also possible to adapt the Cartesian grid to the molecular structure or to perform symmetry transformations on it.

During an \orbkit{} calculation, a LOG file is written that contains the basic information concerning the selected computational parameters, e.g., grid type, grid parameter, chosen input or output file, or the progress of the calculation.
Subsequently, the results, i.e., the electron density, the MOs, etc., given on the user-defined grid, are typically saved as HDF5 files.\cite{hdf5}
This hierarchical data format can efficiently store and organize numerical data with a small need of disk space. Furthermore, there are many programs and tools that support this data type.

For 3D visualizations with standard molecular graphics programs such as VMD\cite{vmd}, \orbkit{} provides the option to save the data as Gaussian Cube files. 
These plain text files contain the volumetric data, the grid parameters, and the atomic positions of the molecular system. 
Besides, \orbkit{} can create VMD script files, which are directly callable with VMD for a quick depiction of any position space function.
It is also possible to use a simple interface to Mayavi\cite{mayavi}, which enables an immediate and interactive visualization.
In addition to the output of grid-dependent quantities, there exists the possibility to create XYZ and PDB \cite{pdb} files with \orbkit{}.

\subsection*{\sffamily \large Features}

Apart from the usual wavefunction analysis (cf. ``Standard Quantities'' in Tab. \ref{tab:features}), several quantum chemical outputs can be compared simultaneously to highlight for instance the influence of the change in the nuclear positions on the electronic structure.
In this context, the ordering routine of \orbkit{} should be mentioned, which enables the correct arrangement of the MOs for different nuclear configurations according to their overlap (cf. Eq. \ref{moom}).
This procedure may be useful when the MOs change their symmetry and/or energetic ordering with the nuclear configuration. 
An example for the usage is given below.
Another valuable feature of \orbkit{} is the adaptive integration of multidimensional functions using a Python wrapper\cite{pycubature} for the C package Cubature\cite{cubature}.
This module integrates numerically vector-valued integrands over hypercubes.
In the implemented $h$-adaptive integration scheme\cite{pub1_cubature,pub2_cubature}, the integrands are iteratively evaluated by gradually adding more grid points until a user-defined tolerance is fulfilled.
Especially for functions with localized sharp features such as the one-electron density, this is a well-suited technique.
Additionally, Cubature enables a simultaneous calculation of the integrals of multiple functions, e.g., MO densities.

\section*{\sffamily \Large Practical Applications}

In this section, we present applications to five molecules: benzene C$_6$H$_6$, 2-cyclopropenyl cation (C$_3$H$_3$)$^+$, hydrogen molecule ion H${}_2^+$, carbon dioxide CO$_2$, and formaldehyde CH$_2$O.
For all examples, the Molden data file and the execution command or an extensively commented Python execution code are provided in the \orbkit{} package.
Thus, the reader is encouraged to follow these examples interactively.
Additional example codes are also included in the \orbkit{} package.
All electronic structure calculations are performed with Molpro\cite{molpro} using the Hartree-Fock method and a cc-pVDZ basis set.\cite{dunning_bs}
There are also some applications of \orbkit{} in the literature, see Refs. \cite{hermann_pohl_perez,Gomez_2015,gesi,bobs}.

\subsection*{\sffamily \large Electron Density of Benzene}

\begin{figure}
\begin{center}
  (a)\includegraphics[width=0.25\columnwidth,keepaspectratio=true]{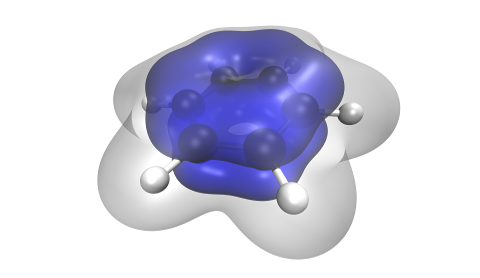}
  (b)\includegraphics[width=0.25\columnwidth,keepaspectratio=true]{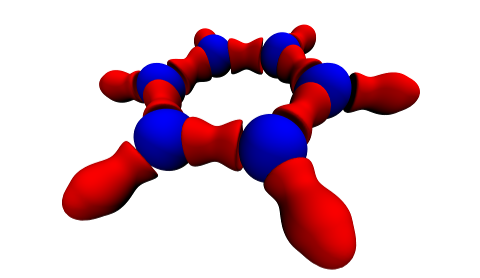}
  (c)\includegraphics[width=0.25\columnwidth,keepaspectratio=true]{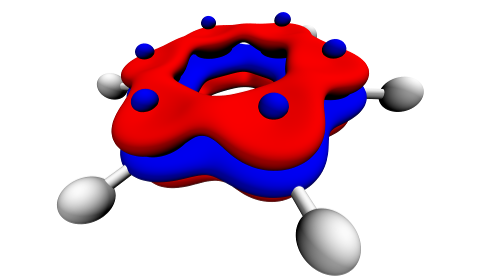}
\end{center}
\caption{\label{benzene} (a) The electron densities of benzene for all electrons (gray) and for the $\pi$-electrons (blue). The isocontour value for the electron densities is 0.01 $a_{0}^{-3}$. 
(b) Laplacians of the molecular electron density and (c) of the $\pi$-electron density for benzene. The isocontour values for the Laplacian of the total electron density are -0.5 $a_{0}^{-5}$ (red) and 0.5 $a_{0}^{-5}$ (blue) 
and -0.1 $a_{0}^{-5}$ (red) and 0.1 $a_{0}^{-5}$ (blue) for the $\pi$-electron density. The visualization was performed with VMD. 
}
\end{figure}

In the first application, the usage of \orbkit{} as a standalone program via the terminal interface and the subsequent visualization of grid-based one-electron quantities are  demonstrated with the example of benzene.
In order to characterize the nature of a chemical bond in a given quantum system, the calculation and analysis of the electron density and its Laplacians, as well as the partitioning of the electron density in certain subsets of electrons, are useful tools.\cite{benzene_schild,laplacian_bo,laplacian_bo2}
In the benzene molecule, for example, we can look at the $\pi$-electron density.
The respective MOs can be identified by their nodal plane being the plane spanned by the nuclei.
Hence, the associated $\pi$-electron density is distributed above and below the benzene ring.
The electron density $\rho_{e} \left( \mathbf{r}\right)$ (cf. Eq. \ref{el_dens}) for all electrons and for the specified set of electrons ($\pi$-electrons) and the respective Laplacians $\nabla^{2} \rho_{e} \left( \mathbf{r}\right)$ (cf. Eq. \ref{laplacian}) are calculated with \orbkit{} and visualized (cf. Fig. \ref{benzene}) with VMD\cite{vmd}.

The Laplacian of the total electron density is depicted in Fig. \ref{benzene}(b). 
Here, the negative Laplacian between two bonded carbons indicates a local concentration of electron density and thus, as expected, an attractive covalent character for this homonuclear bond.
From the Laplacian of the $\pi$-electron density, an accumulation of electron density above and below the ring and a depletion in the plane are noticeable.

Tasks such as the selection of groups of MOs and the subsequent calculation of grid-based one-electron quantities are straightforward with \orbkit{}. 
The associated data can be stored in output formats like Gaussian cube files.
This facilitates its visualization with graphical programs such as VMD, etc. 
In addition, a direct visualization in \orbkit{} is feasible also with Python packages, e.g., matplotlib\cite{matplotlib} or Mayavi\cite{mayavi}.

\subsection*{\sffamily \large Angular Electron Density of (C$_3$H$_3$)$^+$}

\begin{figure}
\begin{center}
  \includegraphics[keepaspectratio=true]{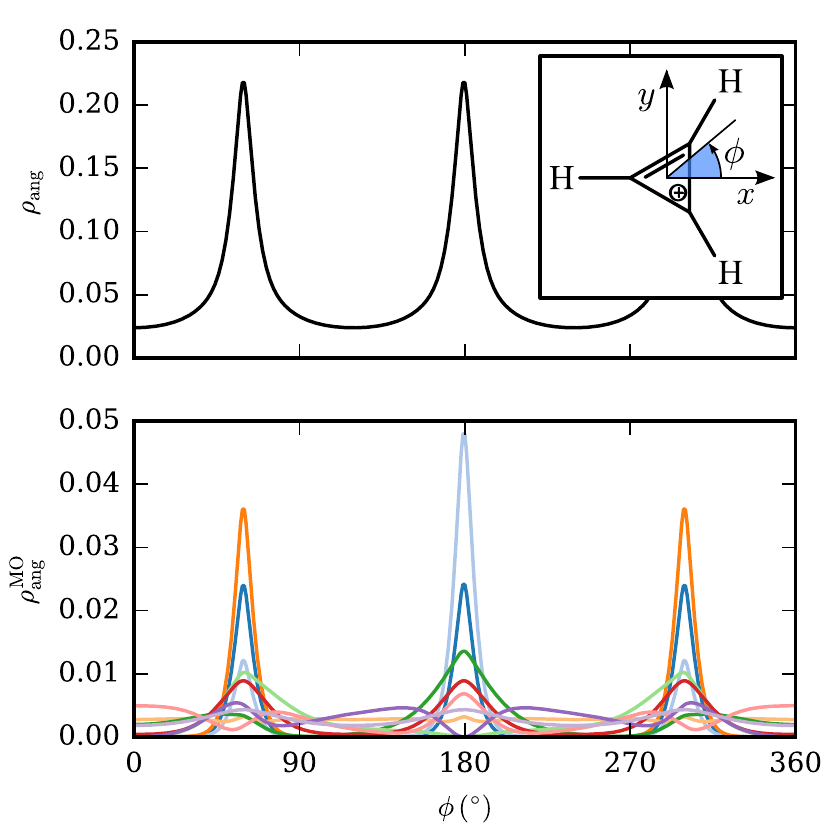}
\end{center}
\caption{\label{angular_int} Angular electron density $\rho_{\rm ang}$ (upper panel) and integrated molecular orbital densities $\rho_{\rm ang}^{\rm MO}$ (lower panel) for angular segments of $\Delta \phi=1^{\circ}$ for the 2-cyclopropenyl cation (C$_3$H$_3$)$^+$ using a Cubature interface. The inset in the upper panel shows the Lewis structure of the 2-cyclopropenyl cation and the orientation of the polar angle $\phi$.}
\end{figure}

Our second application illustrates the ease of utilizing the \orbkit{} library with an existing program.
In this case, the interface of \orbkit{} to Cubature\cite{pub1_cubature,pub2_cubature} is introduced and its virtue for integrating the density in a region of space is demonstrated.
The example at hand shows the integration of the three-dimensional electron density
$\rho(x,y,z)$ of (C$_3$H$_3$)$^+$ to obtain the angular electron density
$\rho_{\rm ang}(\phi)$. 
The angle $\phi$ is defined as the polar angle in the plane of the nuclei (cf.\ inset of Fig. \ref{angular_int}). 
Thus, an integration over $z$ and over $r =\sqrt{x^2+y^2}$ has to be performed. Additionally, to obtain a smooth
angular density close to the positions of the nuclei, an averaging
along $\phi$ has to be done. Hence, for each discrete point $\phi_i$
on our grid we have to integrate over $z \in [-\infty,\infty]$,
$r \in [0,\infty]$, and $\phi \in [\phi_i-\Delta \phi/2,
\phi_i+\Delta \phi/2]$.

The choice of grid coordinates is not as straightforward as it may seem.
While cylindrical coordinates are a natural choice, they have a major
disadvantage if used equidistantly: Because of the fixed number of points along $\phi$, there will be a high density of points for small $r$ but very few
for larger $r$. Thus, the number of points needed to accurately integrate the density
around the nuclei quickly becomes prohibitively large.
Calculating the density on a grid in Cartesian coordinates is not an alternative.
While the point density in space does not change, the number of points per angle varies significantly. 
As a consequence, integration using Cartesian coordinates needs far too many points to calculate the angular density efficiently and can yields artifacts nevertheless (see chapter 3.4 of  Ref. \cite{thesis_schild} for an example).

Thus, we implemented an interface to a Python wrapper\cite{pycubature} for Cubature\cite{cubature}.
This program can integrate multidimensional functions with moderate dimensionality adaptively to a specified error.
A function for integrating in cylindrical coordinates converts the points asked for by Cubature (which are assumed to be cylindrical coordinates $r, \phi, z$) into Cartesian coordinates, calls \orbkit{} using the list of Cartesian vectors as input, and returns the  density multiplied with $r$ to account for the volume element of the integration. 
This implementation is straightforward, as \orbkit{} is designed to be used as both a standalone program and a function library.

Figures \ref{angular_int}(a) and (b) show the angular density and the integrated molecular
orbital densities, respectively. The integration of the regions of strong localization
around the nuclear positions are well-converged, which would have been difficult
to achieve using simple integration schemes on equidistant grids.

In general, the adaptive integration by Cubature of grid-based quantities computed in \orbkit{}, in any user-defined volume, opens up a wide spectrum of conceivable applications, for example, the determination of Voronoi deformation density atomic charges, etc.

\subsection*{\sffamily \large Ordering Molecular Orbitals Along a Reaction Coordinate}

In quantum chemistry, it can be of interest to follow the change of molecular properties during the variation of the nuclear structure, for example, along a reaction coordinate.
While the comparison of most properties is straightforward but possibly cumbersome because data may have to be extracted from several output files, the comparative analysis of the molecular orbitals can be a problem:
Most quantum chemistry programs are sorting the orbitals according to their energy and additionally, if applicable, according to their symmetry.
However, often the energetic order of the orbitals changes with the nuclear configuration, and it is necessary to order them according to a different criterion.
For these issues, \orbkit{} offers a set of useful functions to simultaneously handle multiple files, and it possesses several MO ordering routines.

The most useful ordering routine sorts the MOs of different nuclear configurations and the associated signs according to their overlap.
It starts from the first two structures in a list, computes the analytical overlap between all the orbitals (cf. Eq. \ref{moom}) of both structures, and sorts them accordingly.
Thereafter, it proceeds with the second and third structure, etc.
Subsequently, the expansion coefficients for the ordered orbitals data can be interpolated using B-Splines to approximate intermediate nuclear configurations.
B-Splines (cf. Ref. \cite{bsplines} and references therein) are piecewise polynomial functions with many useful properties,
one of which is their analytical differentiability. This substantiates their usage for the accurate determination of non-adiabatic coupling terms.

\begin{figure}
\begin{center}
\includegraphics[keepaspectratio=true]{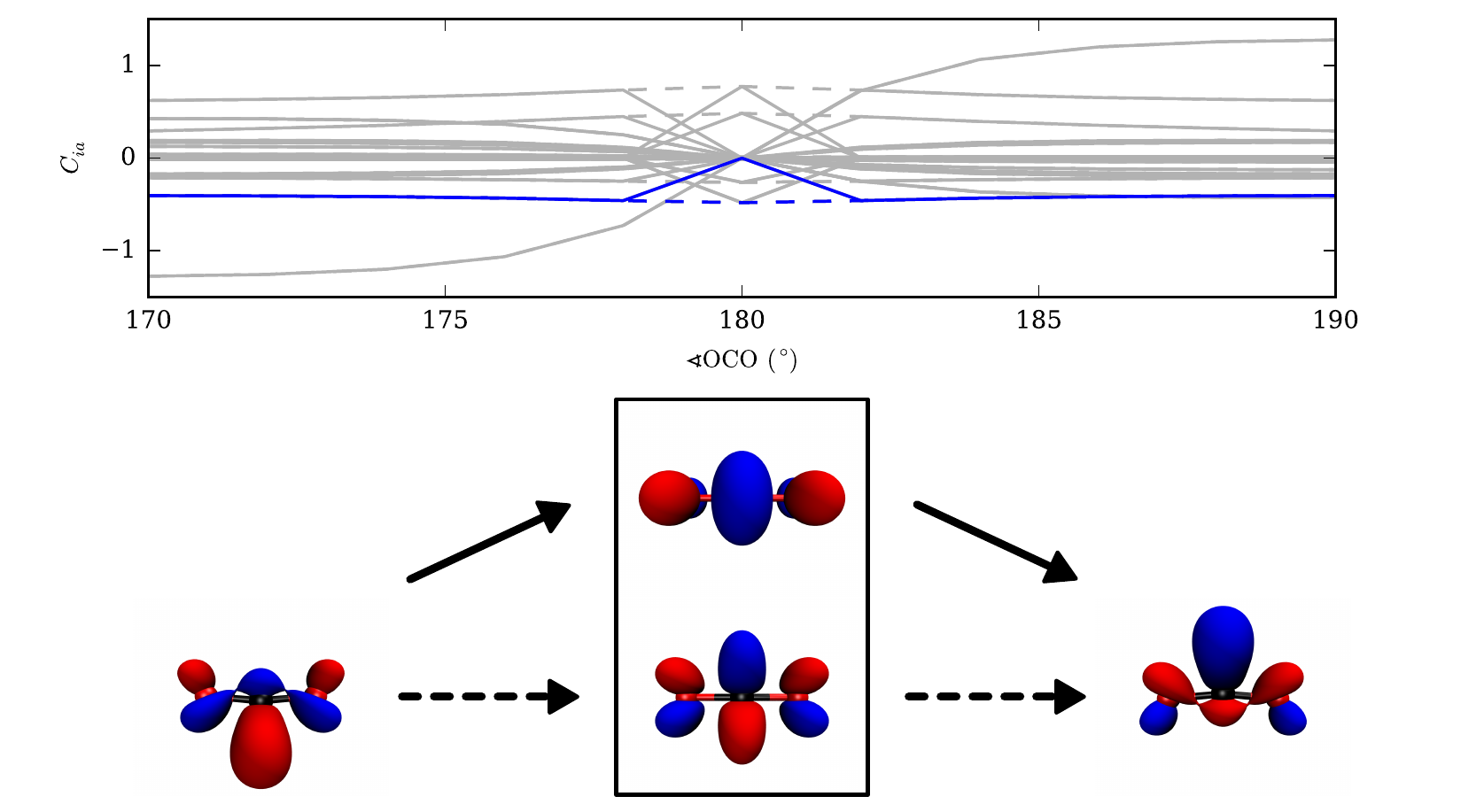}
\end{center}
\caption{\label{mo_order} 
Upper panel: Molecular orbital coefficients $C_{ia}$ of the MO (initially the LUMO) as a function of the $\sphericalangle$ OCO angle before (solid line) and after (dashed line) sorting by \orbkit{}.
Lower panel: Isosurface plots of the lowest unoccupied molecular orbital (LUMO) of CO$_2$ at an O-C-O angle of 170$^\circ$ (left), of the LUMO and LUMO+1 (degenerate) at 180$^\circ$, and of the LUMO at 190$^\circ$. Solid and dashed arrows correspond to the solid and dashed lines in the upper panel. The isosurface value is $\pm$0.1 $a_{0}^{-3}$. The isosurface plots were visualized with VMD.}
\end{figure}

To illustrate the functionality of this ordering routine, we performed a set of quantum chemical calculations for CO$_2$, varying the $\sphericalangle$ OCO angle from 170$^\circ$ to 190$^\circ$ with a step size of 2$^\circ$.
For a sorted set of molecular orbitals, one expects smooth curves for the coefficients as a function of the slightly modified nuclear configurations.
In Fig. \ref{mo_order} (upper panel), the MO coefficients are displayed for a selected orbital of CO$_2$. Solid lines correspond to the MO coefficients $C_{ia}$ for the lowest unoccupied MO (LUMO) of the energetically ordered list of the quantum chemical output and dashed lines signify the coefficients of the orbital after sorting them according to the MO overlap. To illustrate the difference between a sorted and unsorted MO list, the curve of one chosen coefficient is marked in blue.
For linear CO$_2$  ($\sphericalangle {\rm OCO=180^\circ}$), the selected orbital, the LUMO, is energetically degenerate with another one, the LUMO+1, which leads to an interchange of both in the energetically ordered list of the quantum chemistry program.
Following the procedure described above, \orbkit{} can sort all orbitals according to their overlap and in groups according to their symmetry properties, if this information is available.
The lower panel of Fig. \ref{mo_order} shows the orbitals that were incorrectly assigned to each other (solid arrows) and those assigned to each other after ordering (dashed arrows).
Note that the results of the ordering routine depend on the validity of the overlap as a measure of the character of the orbital. 
Hence, the prerequisites for a successful MO sorting are moderate structural variations between the quantum chemical calculations and the identical orientation of the molecule.
Nonetheless, a failure of the ordering routine can be easily detected and corrected by inspecting the graphs of, e.g., the orbitals energies or orbital coefficients, and by using the manual ordering function of \orbkit{}.
To the best of our knowledge \orbkit{} is the only post-processing program that provides such an orbital ordering function.
This is complemented by a number of convenience functions to plot, save and load the computed quantities.

\subsection*{\sffamily \large Transition Electronic Flux Density of H${}_2^+$}

The example of this section illustrates the computation of the stationary transition electronic flux density (TEFD) between two selected electronic Born-Oppenheimer states in the hydrogen molecular ion  H$_2^+$ with \orbkit{}.\cite{hermann_pohl_perez} 
In general, the TEFD is the non-vanishing component of the electronic flux density in the framework of the Born-Huang expansion and is defined for the transition from the electronic state $\lambda$ to the electronic state $\nu$ as
\begin{eqnarray}
  \mathbf{J}_{e,\lambda \nu}^{\rm TEFD} (\mathbf{r},t) & = & \int {\rm d}\mathbf{R} ~ \rho_{{n,\lambda \nu}}(\mathbf{R},t) \cdot \mathbf{J}_{{e,\lambda \nu}}^{\rm{STEFD}} (\mathbf{r};\mathbf{R}),
\end{eqnarray}
with the nuclear transition density $\rho_{{n,\lambda \nu}}(\mathbf{R},t) = \chi_\lambda^\star(\mathbf{R},t) \chi_\nu(\mathbf{R},t)$ and the purely imaginary static transition electronic flux  density (STEFD)
\begin{eqnarray}
  \mathbf{J}_{{e,\lambda \nu}}^{\rm{STEFD}} (\mathbf{r};\mathbf{R}) &=& -\frac{\imath}{2} \left(\Psi_\lambda(\mathbf{r};\mathbf{R}) \nabla_e \Psi_\nu(\mathbf{r};\mathbf{R}) - \Psi_\nu(\mathbf{r};\mathbf{R}) \vec{\nabla}_e \Psi_\lambda(\mathbf{r};\mathbf{R}) \right)\,,
\end{eqnarray}
where $\Psi_\lambda, \Psi_\nu$ are the real-valued electronic wavefunctions and the gradient $\vec{\nabla}_e$ is taken with respect to the electronic coordinates.\cite{nafie1997}
The TEFD is a crucial quantity to analyze the contributions to infrared absorption or vibrational circular dichroism spectra\cite{freedman97,nafie11}, as a complement to the study of the adiabatic electronic flux density\cite{hermann_pohl_perez}, or for the visualization of electron processes, for example.

The hydrogen molecular ion H$_2^+$ is the simplest diatomic molecule consisting of two protons and one electron. 
The fact that the molecular orbitals of H$_2^+$ correspond to its electronic states simplifies the calculations of the TEFD.
However, it is nonetheless feasible to determine this quantity for more complicated quantum systems with \orbkit{}.
This can be accomplished with the help of the Slater-Condon rules, but 
it is necessary to take into account the underlying basis set expansion of the wavefunction (e.g.\ single Slater determinant, 
configuration state functions, multireference configuration interaction representation, etc.).
The required modules for this computational task can be easily incorporated into the modular structure of \orbkit{}.
In addition, the simple and platform-independent parallelization techniques within Python can be used to enhance the efficiency of the implemented code.
Recently, \orbkit{} was used to investigate the ultrafast photoelectron transfer in a Dye-Sensitized Solar Cell by analyzing the corresponding time-dependent TEFD constructed from configuration interaction wavefunctions.\citep{Gomez_2015}

\begin{figure}
\begin{center}
\includegraphics[keepaspectratio=true]{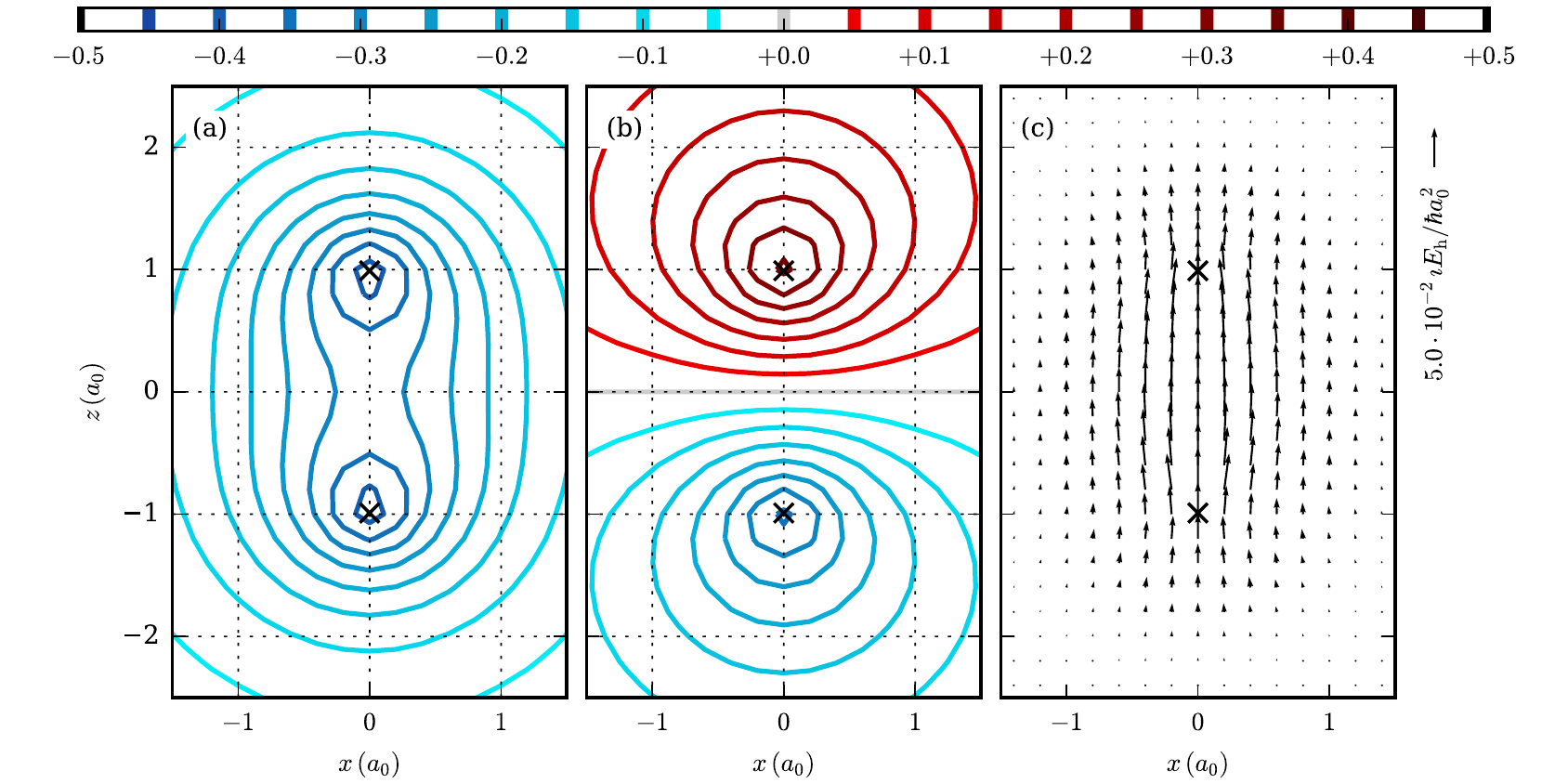}
\end{center}
\caption{\label{h2+} Contour plots of selected molecular orbitals (MO) of the hydrogen molecule ion H${}_2^+$ for an internuclear distance of $R={1.4\, a_0}$: (a) MO 1$\sigma_{g}$ and (b) MO 1$\sigma_{u}$. (c) Vector plot of the stationary transition electronic flux density (STEFD) $\mathbf{J}_{{e,1\sigma_g 1\sigma_u}}^{\rm{STEFD}}$ for the transition between the state 1$\sigma_{g}$ and the state 1$\sigma_{u}$.
The nuclear positions are marked with black crosses.}
\end{figure}

For the TEFD of H$_2^+$, the transition between the electronic ground state $1\sigma_g$ and the first excited state $1\sigma_u$ is selected, since it is experimentally accessible.\cite{Bolognesi}
Contour plots of both electronic states (MOs) are displayed in Fig. \ref{h2+}(a) and \ref{h2+}(b) showing their gerade and ungerade symmetry properties. 
The stationary TEFD $ \mathbf{J}_{{e,1\sigma_g 1\sigma_u}}^{\rm{STEFD}}$ for the transition between the ground state $1\sigma_g$ and the excited state $1\sigma_u$ as a function of the $x$- and $z$-coordinate for an internuclear distance of $R=1.4\,{a_0}$ is depicted in Fig. \ref{h2+}(c). 
As expected, this imaginary vector field shows a gerade parity for the transition between a gerade and an ungerade state.

\subsection*{\sffamily \large Partial Charges and Gross Atomic Density}

\begin{figure}
\begin{center}
\includegraphics[width=0.3\columnwidth,keepaspectratio=true]{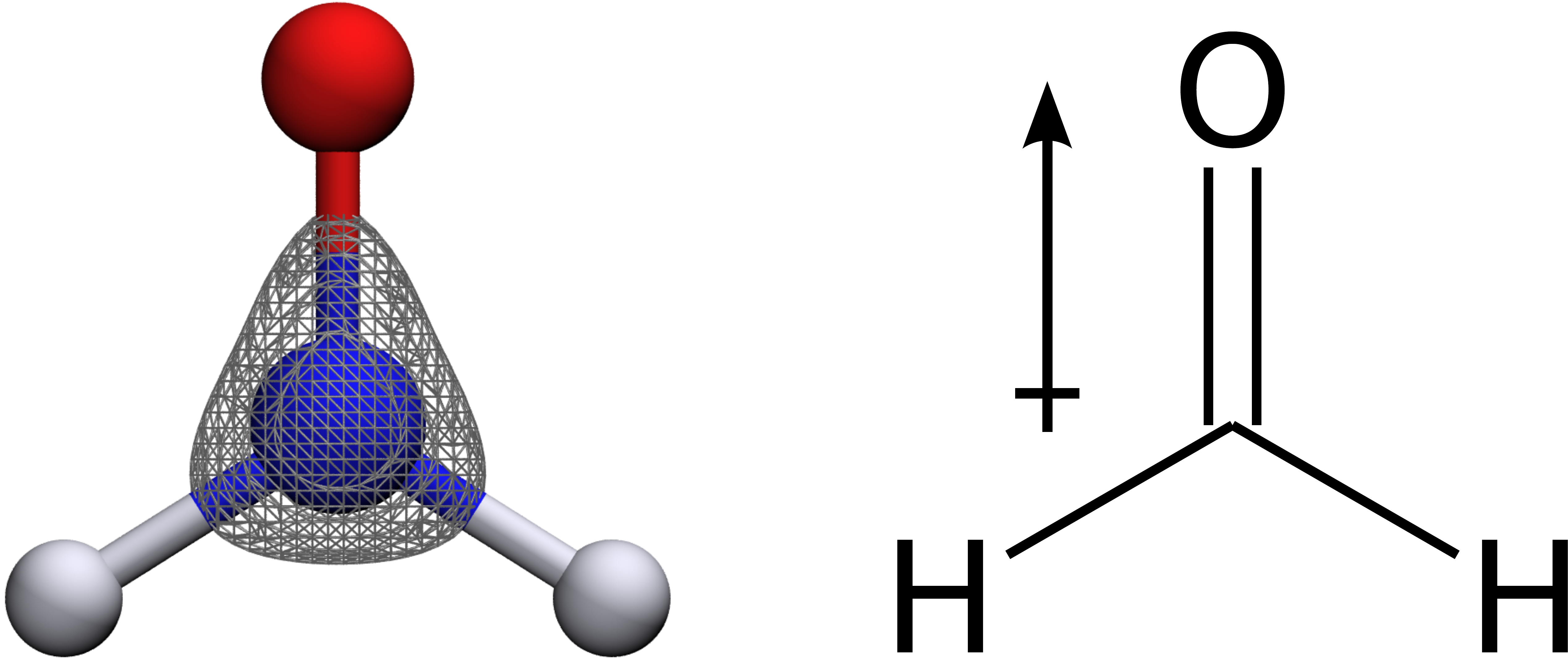}
\end{center}
\caption{\label{ch2o} Ball-and-stick representation and Lewis structure of a formaldehyde molecule. Balls and sticks are colored according to the Mulliken partial charge of the respective atom. Positive partial charges are blue, and negative partial charges are red. The direction of the molecular electric dipole moment is shown as an arrow next to the Lewis structure. The gross atomic density of the carbon atom is represented as a gray wireframe with isocontour value 0.2 $a_{0}^{-3}$. The ball-and-stick representation and the wireframe plot were illustrated with VMD\cite{vmd}.
}
\end{figure}

\orbkit{} possesses a variety of specialized modules to determine various properties of a molecular system.
In the present example, several of these modules are used for the formaldehyde molecule.
To start with, we compute two non-grid based quantities: the Mulliken partial charges\cite{jensen}
\begin{eqnarray}
q_{A} = Z_{A} - \sum_{i\in A}^{N_{\rm AO}} \sum_{j}^{N_{\rm AO}} \left( \sum_{a}^{N_{\rm{occ}}}n_aC_{ia}C_{ja} \right) \braket{ \psi_{i}|\psi_{j}}
\end{eqnarray}
and the total electric dipole moment\cite{jensen}
\begin{eqnarray}
\boldsymbol{\mu} = \Braket{\varphi_{a}\left|-\sum_{a}^{N_{\rm{occ}}}\mathbf{r}_{a} \right|\varphi_{a}}+ \sum_{A} Z_{A}\mathbf{R}_{A}.
\end{eqnarray}
To visualize partial charges on a molecular system,  \orbkit{} offers the opportunity to create PDB files containing the type, position, and the partial charges of the nuclei.
A variety of visualization programs has the ability to depict PDB files. In Fig. \ref{ch2o},
we use a ball-and-stick representation colored according to the partial charges, i.e., red for negative charges and blue for positive charges.
Additionally, the molecular electric dipole moment of formaldehyde is represented as a position vector in the adjacent Lewis structure.

A related grid-based quantity to the Mulliken partial charges is the gross atomic density\cite{mulliken}
\begin{eqnarray}
\rho_{A}\left(\mathbf{r}\right) = \sum_{i\in A}^{N_{\rm AO}} \sum_{j}^{N_{\rm AO}} \left( \sum_{a}^{N_{\rm{occ}}}n_aC_{ia}C_{ja} \right) \psi_{i}\left(\mathbf{r}\right)\psi_{j}\left(\mathbf{r}\right)
\end{eqnarray}
whose integral coincides with the respective gross atomic population.
In Fig. \ref{ch2o}, the gross atomic density for the carbon atom is shown as a gray wireframe.

Gross atomic populations and Mulliken partial charges are standard quantities, which can be computed by many other programs as well. 
\orbkit{} provides the feature to compute them on a grid to enable a simple consistency check of the grid convergence.
Once the value for the integral over the Gross atomic density converges to the Gross atomic population analytically calculated, the grid size is appropriate for the evaluation of integrated quantities.

\section*{\sffamily \Large Conclusion}

\orbkit{} is a modular designed Python toolbox that allows an individualized cross-platform post-processing of quantum chemical data from electronic structure calculations.
The variety of position space one-electron functions and fundamental quantities that has been implemented serves as the basis for sophisticated analyses of  molecular wavefunctions. 
Thus, it is useful for a wide range of applications.
In addition, in its current state of development \orbkit{} offers multiple options and features for post-processing issues.
For the calculation of one-electron quantities on arbitrary grids, there exists a standalone version which is easy to handle and can be executed in parallel to speed up the computation.
The results can be directly visualized with a simple and interactive viewer (Mayavi).
This is complemented by the interoperability of \orbkit{} with various external visualization programs.
Besides the standalone execution, the functions existing in \orbkit{} can be individually combined to enable a problem-specific wavefunction analysis.
Furthermore, the possibility to add user-written Python functions into \orbkit{} can foster the development of own post-processing programs.
For this purpose, only a basic understanding of the design and features of \orbkit{} is required.
The first steps into the program are facilitated by a detailed documentation and several application examples.
Hence, \orbkit{} appeals to novices as well as experienced theoretical chemists.\\
As a wavefunction analysis toolkit, \orbkit{} differs from similar projects by its portability and its simple modular structures.
They enable the user to build own application programs on top of \orbkit{} with minimal effort and without recompiling.
In conclusion, \orbkit{} is a fairly mature open-source program that provides the basis for many possible further developments.

\subsection*{\sffamily \large Acknowledgments}

The authors thank the Scientific Computing
Services Unit of the Zentraleinrichtung f\"ur Datenverarbeitung
at Freie Universit\"at Berlin for allocation of computer time.
J.C.T. and G.H. acknowledge the funding of the Deutsche Forschungsgemeinschaft (DFG)
through the Emmy-Noether program (project TR1109/2-1) and V.P. of the Elsa-Neumann foundation of the Land Berlin.
Furthermore, G.H., V.P.\ and A.S.\ acknowledge funding by the DFG within the grant Ma 515/25-1.

\clearpage


\bibliography{orbkit.bib}   


\clearpage

\end{document}